\documentclass{PoS}
\usepackage{verbatim}
\usepackage{subfigure}

\title{B and bottomonium spectroscopy
from lattice NRQCD with charm
in the sea
}

\ShortTitle{Bottomonium spectroscopy from lattice NRQCD}

\author{
\speaker{R.~J.~Dowdall}$^a$
for the HPQCD Collaboration
\\ \\
   \llap{$^a$}School of Physics and Astronomy, University of Glasgow, Glasgow, UK\\
\\
        E-mail: \email{Rachel.Dowdall@glasgow.ac.uk}
}

\abstract{
We give results for $B$, $B_s$, $B_c$ and bottomonium spectroscopy using NRQCD heavy quarks and HISQ valence and sea quarks. Five MILC ensembles of gluon configurations with three values of the lattice spacing and $m_{\rm light}/m_{\rm strange}$ values down to 0.1 are used that include 2+1+1 flavours of sea quark. Systematic errors in the NRQCD action are improved through the radiative correction of the coefficients of terms at $v^4$. Improved results for S-wave and P-wave bottomonium states are discussed as well as a prediction for the full D-wave spectrum. 
Preliminary results for the ground state B meson masses are also presented.
}

\FullConference{ The XXIX International Symposium on Lattice Field Theory - Lattice 2011\\
July 10-16, 2011\\
Squaw Valley, Lake Tahoe, California}

\begin{document}

\section{Introduction}
The spectroscopy of mesons containing $b$ quarks has been an important testing ground for lattice QCD \cite{Gray:2005ur} and 
recent improvements in a number of areas mean it is necessary to revisit the bottomonium and B meson spectra.
The study of the bottomonium spectrum on the lattice is simpler than for lighter hadrons in several important ways.
There are a number of states below the threshold for decay into two B mesons meaning that for low lying states there is no systematic error from not including these multiparticle diagrams on the lattice. The splittings between excited states in the spectrum show little dependence on the valence $b$ quark mass, which reduces the error due to mistuning, or the light sea quark masses, which reduces errors from chiral extrapolation.

The use of effective field theories such as NRQCD for $b$ quarks avoids the large discretisation errors that are typical of other lattice quark formulations on coarse lattices. NRQCD is also computationally very cheap since propagators can be calculated from a time evolution rather than an inversion of the Dirac matrix.
In recent years the HPQCD collaboration have calculated order $\alpha_s$ radiative corrections to the Wilson coefficients in the NRQCD action which were the main source of systematic error. We present a new study of the bottomonium spectrum, along with preliminary results for the lowest lying B meson states, using these coefficients and a number of other improvements over the previous HPQCD calculation. This serves as a check of the improvements made and allows the remaining parameters, such as the mass and lattice spacing, to be fixed. We are also able to make a prediction for the D-wave states.
Successfully reproducing the known part of the spectrum gives us confidence in more phenomenologically interesting quantities, such as decay constants and mixing matrix elements, which will follow on from this calculation.

Further improvements have been made in the gluon and sea quark sector. 
We use the recent MILC collaboration ensembles with  2+1+1 flavours of HISQ sea quarks at three lattice spacings (from 0.15fm to 0.09fm) and two light quark masses ($m_l/m_s = 0.1,0.2$) \cite{Bazavov:2010ru}. 
These have more chiral light quarks, include charm for the first time and have a larger spatial volume than previous ASQTAD ensembles.
The coefficients of the gluon action have also been perturbatively improved \cite{Hart:2008sq}, including the effect of the HISQ sea quarks.

Full details of the S and P-wave results presented in these proceedings can be found in \cite{upsilonpaper}.

\section{Calculation details}

\paragraph{Heavy quark action:}
NRQCD is an expansion of QCD in powers of the heavy quark velocity $v$.
The form of the NRQCD Hamiltonian is:
\begin{eqnarray}
aH_0 &=& - \frac{\Delta^{(2)}}{2 am_b} \nonumber \\
a\delta H
&=& 
- c_1 \frac{(\Delta^{(2)})^2}{8( am_b)^3}
+ c_2 \frac{i}{8(am_b)^2}\left(\bf{\nabla}\cdot\tilde{\bf{E}}\right. -
	\left.\tilde{\bf{E}}\cdot\bf{\nabla}\right) 
- c_3 \frac{1}{8(am_b)^2} \bf{\sigma}\cdot\left(\tilde{\bf{\nabla}}\times\tilde{\bf{E}}\right. -
\left.\tilde{\bf{E}}\times\tilde{\bf{\nabla}}\right)
\nonumber \\
& & 
- c_4 \frac{1}{2 am_b}\,{\bf{\sigma}}\cdot\tilde{\bf{B}}  
+ c_5 \frac{a^2\Delta^{(4)}}{24 am_b} 
- c_6 \frac{a(\Delta^{(2)})^2}{16n(am_b)^2} \nonumber
\label{deltaH}
\end{eqnarray}
This is the standard action including terms of order $v^4$ and some other terms designed to reduce discretisation errors, see ref. \cite{Lepage:1992tx}. $am_b$ is the $b$ quark mass in lattice units, $n$ is the stability parameter used in the evolution equation, $\tilde{\bf{E}},\tilde{\bf{B}}$ are the improved chromo-electric and chromo-magnetic field strengths and the $\Delta^{(k)}$ are lattice derivatives. These are all defined in more detail in \cite{Gray:2005ur}. 
With the coefficients $c_i=1$, this action is equivalent to QCD at tree level.
The $c_i$ can be expanded in powers of $\alpha_s$, as $c_i = 1 + c_i^{(1)} \alpha_s + \mathcal{O}(\alpha_s^2)$ and most of the $c_i^{(1)}$ have now been calculated perturbatively by HPQCD by matching to QCD at one loop. 
Radiative corrections are kept small by tadpole improving the gauge fields in the action with the Landau link $u_{0L}$.
Tuning of some of the coefficients has also been performed nonperturbatively as an independent check and to estimate higher order contributions.
Radiative corrections are particularly important for quantities such as the hyperfine splitting, which is proportional to $c_4^2$, and were the dominant source of systematic error in previous calculations.

\paragraph{Tuning of parameters: }
The lattice spacings were fixed to a precision of better than $1\%$ using the Upsilon 2S-1S splitting. 
For a consistency check, the static quark potential parameter $r_1$ was also calculated using the decay constant of the $\eta_s$ meson \cite{Davies:2009tsa} and MILC values for $r_1/a$. Both methods give continuum values for $r_1$ that agree.

The $b$ quark mass was tuned by computing the spin averaged kinetic mass 
$
\overline{M_{b \overline{b} } } =  \frac{1}{4}
\left(   
3 M_{\Upsilon} + M_{\eta_b}
\right)
$  and comparing to the experimental value appropriately adjusted for missing electromagnetic and annihilation effects.
In ref \cite{upsilonpaper}, we present a detailed study of systematic errors in the tuning of $am_b$ and provide accurate values for each ensemble. 
The light, strange and charm quarks in the heavy-light calculations were included with the highly improved staggered quark (HISQ) action \cite{Follana:2006rc} and the strange and charm quark masses were tuned to the masses of the $\eta_s$ and $\eta_c$. 
Stochastic noise sources were used for the S and P-wave propagators and all B mesons. Data was extracted from the correlators using a multi-exponential matrix Bayesian fit. Full details of all parameters used are given in \cite{upsilonpaper}.

\section{Results}
\begin{figure}
 \begin{center}
  \begin{minipage}[t]{0.48\hsize}
  \vspace{0pt}
     \includegraphics[width=0.99\hsize]{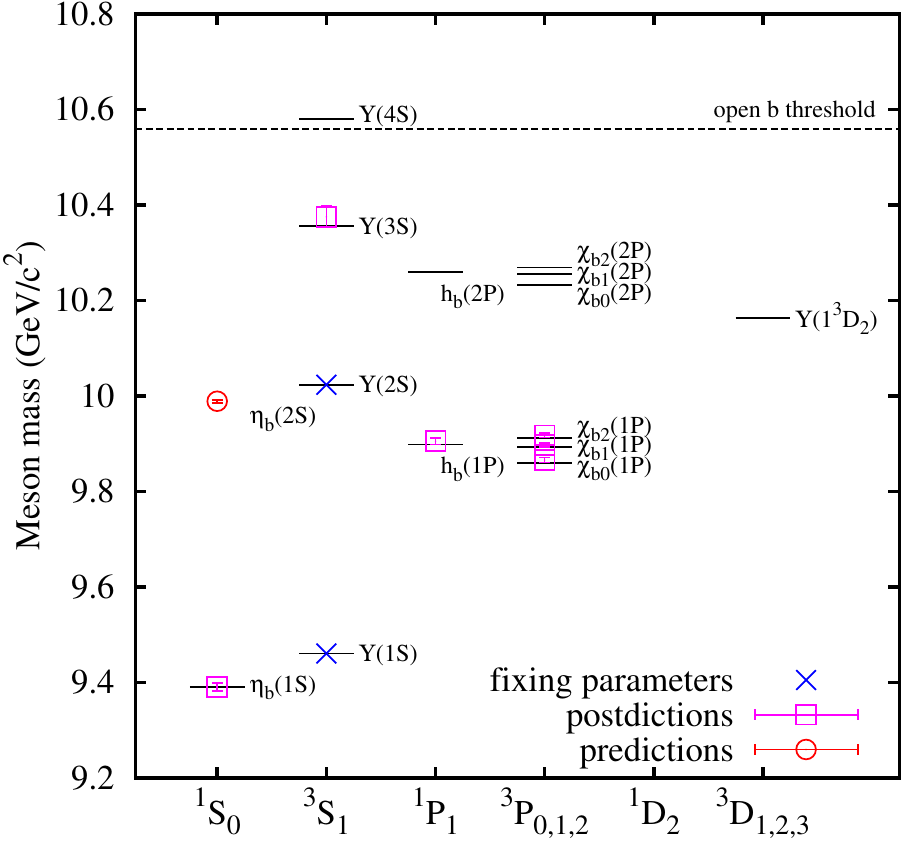}
     \caption{An overview of the spectrum obtained in our calculation. The $\Upsilon(1S,2S)$ are used for fixing the lattice       	spacing and tuning the $b$ quark mass. Other points are pre/post-dictions. }
     \label{fig:spectrum}
  \end{minipage}
  \hfill
  \begin{minipage}[t]{0.48\hsize}
  \vspace{0pt}
   \includegraphics[width=0.95\hsize]{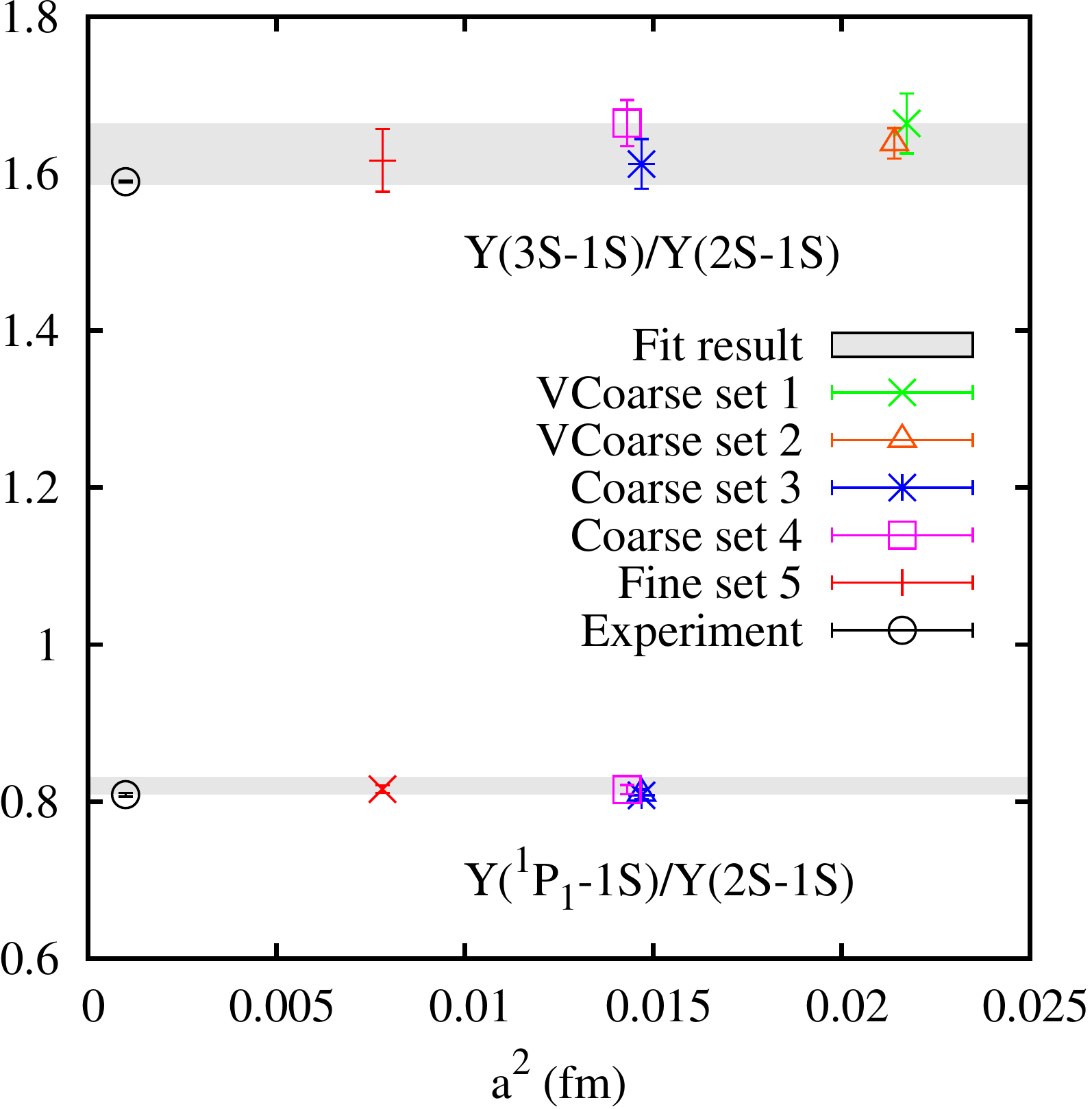}
   \caption {Plot of the ratios of energy splittings for each ensemble along with the physical value obtained from the fit.} 
   \label{fig:ratios}
   \end{minipage}
 \end{center}
\end{figure}

Here we present a subset of the results obtained in \cite{upsilonpaper} starting with an overview of the bottomonium spectrum obtained in figure \ref{fig:spectrum}. 
Not all points on the plot are pre/post-dictions, the Upsilon $2S-1S$ splitting was used to fix the lattice spacing and the spin average of the $1S$ states was used to tune $am_b$.

\paragraph{Splitting ratios:}
The quantities which can be calculated most accurately are ratios of energy splittings. 
Some systematic errors cancel in the ratio leaving an estimated systematic on each ensemble of $< 1\%$. 
Results are plotted in figure \ref{fig:ratios}, and we extract physical values of
\begin{eqnarray}
\frac{ \Upsilon(3S -1S) }{\Upsilon(2S -1S)} = & 1.625(39), & \ \ \ \  
{\rm Expmt} = 1.5896(12) \nonumber \\
\frac{ 1^1P_1 - \overline{1S} }{\Upsilon(2S -1S)}   = & 0.820(12), & \ \ \ \ 
{\rm Expmt} = 0.8088(23) 
\end{eqnarray}
which are in good agreement with experiment (from \cite{PDG}) and have overall errors of $2.4\%$ and $1.4\%$ respectively.

\begin{figure}
 \begin{center}
  \begin{minipage}[t]{0.48\hsize}
  \vspace{0pt}
   \includegraphics[width=0.95\hsize]{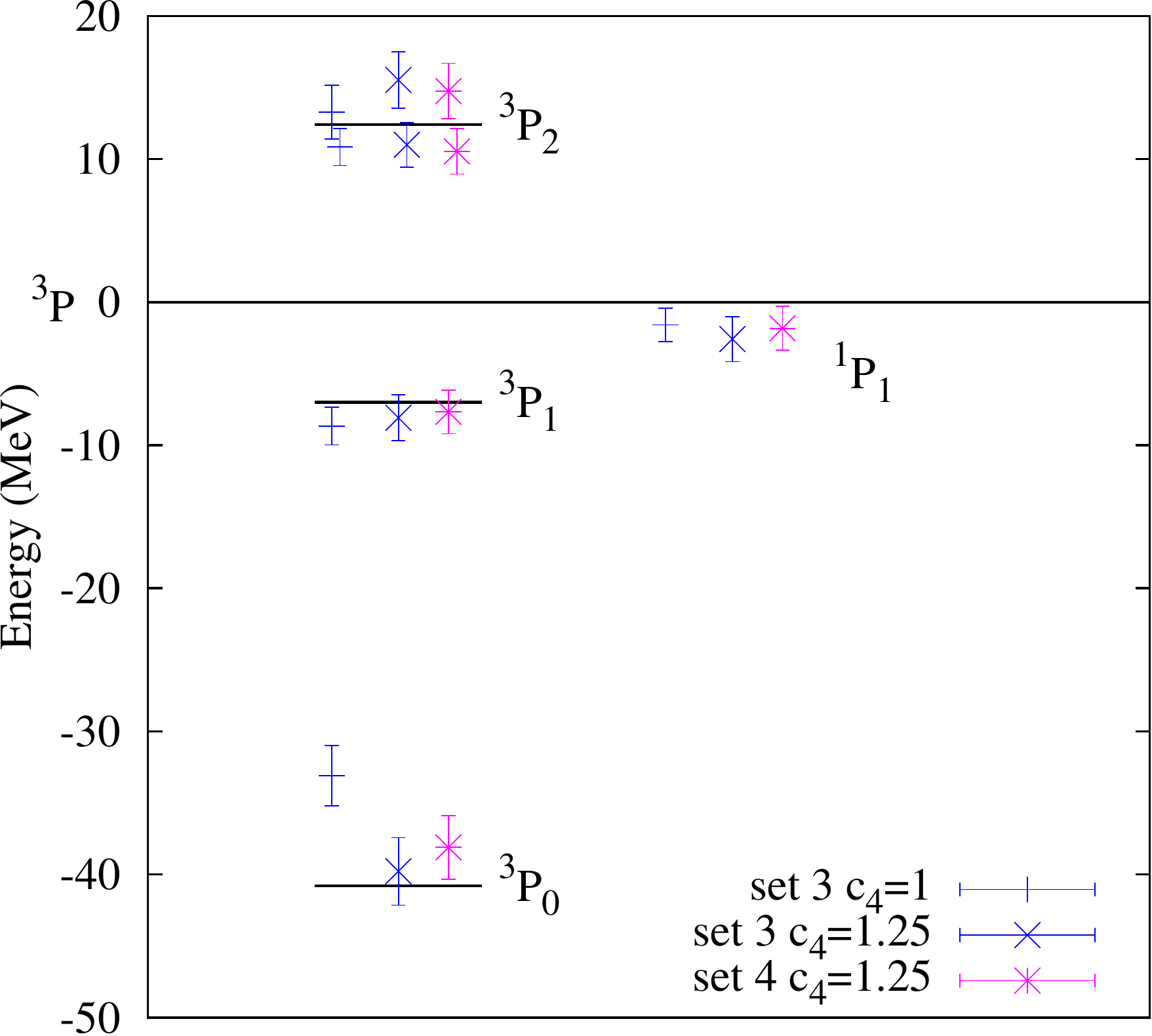}
   \caption{Results for the P-wave spin splittings relative to the spin averaged $^3P$ state. Set 3 and 4 label the two coarse ensembles ($\sim 0.12$fm) in our calculation with $m_l/m_s = 0.2,0.1$ respectively. 
   The two values for the $^3P_2$ state denote the $E$ and $T_2$ lattice irreps (left and right respectively.)}
   \label{fig:pwaves}
  \end{minipage}
  \hfill
  \begin{minipage}[t]{0.48\hsize}
  \vspace{0pt}
   \includegraphics[width=0.95\hsize]{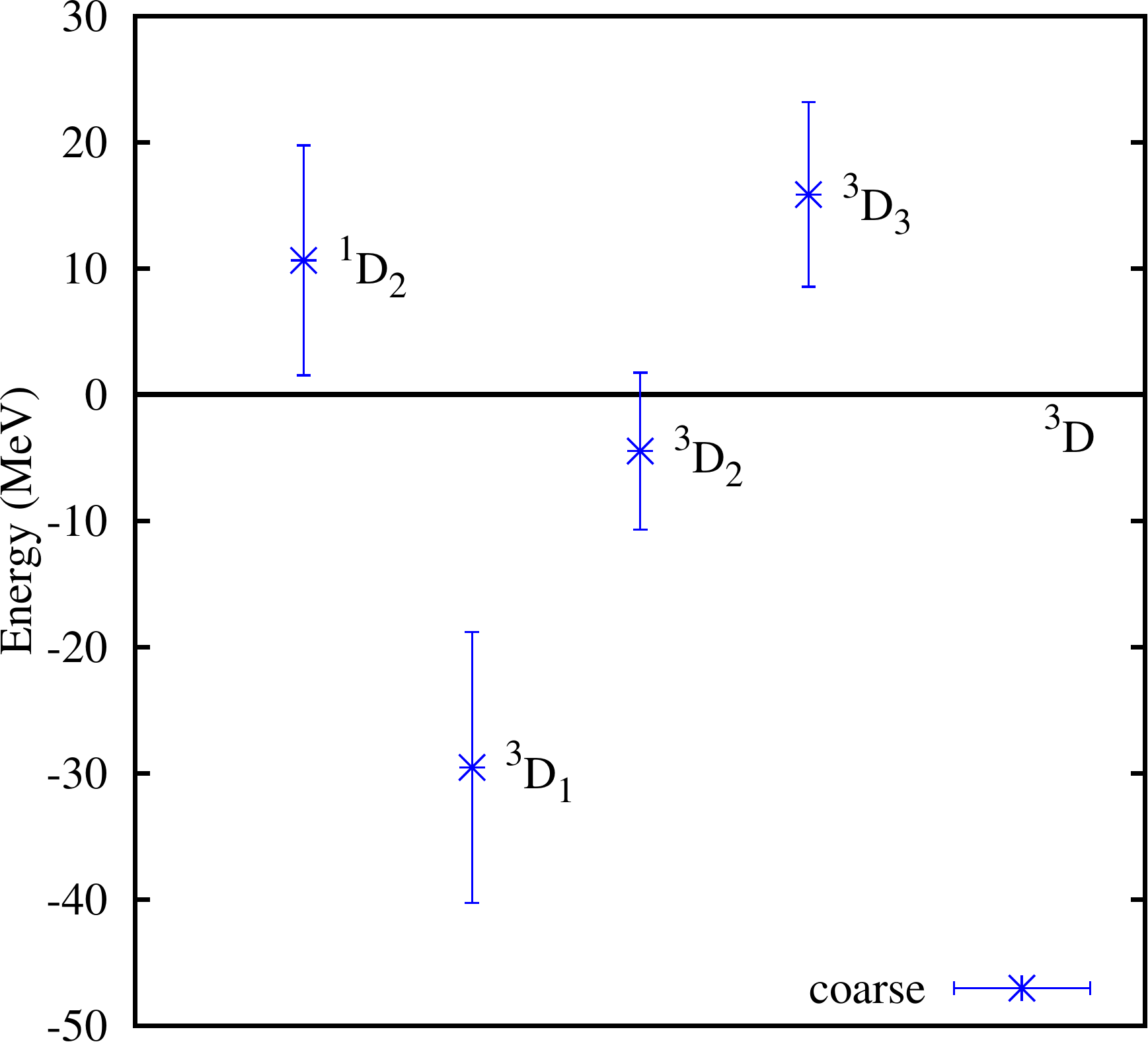}
   \caption{Plot of the D-wave bottomonium states relative to the spin averaged $^3D$ state on a coarse ensemble. Errors are dominated by statistics.}
   \label{fig:dwaves}
  \end{minipage}
 \end{center}
\end{figure}

\paragraph{P-wave splittings and nonperturbative tuning:}
The P-wave spin splittings are shown in figure \ref{fig:pwaves}. 
We tune $c_3,c_4$ (see \cite{Gray:2005ur}) so that these splittings agree with the experimental values (crosses).
The plot also shows the tree level coefficients which give a $^3P_0 - ^3\overline{P}$ splitting that is too small.
This nonperturbatively tuned $c_4$ agrees with the perturbative calculation within errors. The fine lattice P-wave masses are shown on figure \ref{fig:spectrum} for the perturbative value of $c_4$.

\begin{figure}
 \begin{center}
  \begin{minipage}[t]{0.48\hsize}
  \vspace{0pt}
   \includegraphics[width=0.95\hsize]{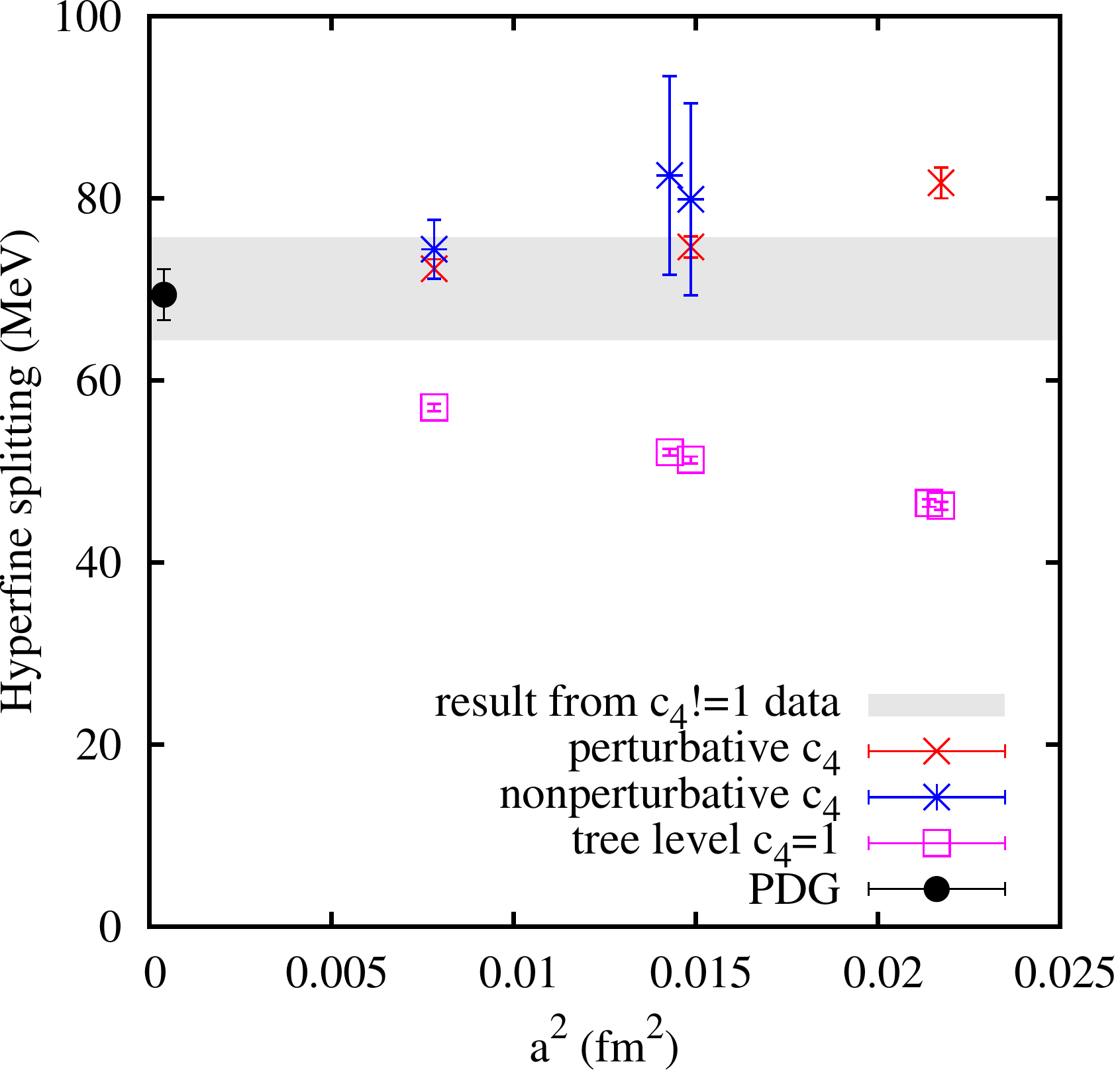}
   \caption{ Results for the hyperfine splitting. Perturbative and nonperturbative $c_4$ values are shown with errors from statistics, lattice spacing, mass retuning and 4-quark operators shown. Correlated higher order systematic errors are not shown on the data. Tree level $c_4=1$ points are also plotted but are not included in the fit}
   \label{fig:hyperfine}
  \end{minipage}
  \hfill
  \begin{minipage}[t]{0.48\hsize}
  \vspace{0pt}
   \includegraphics[width=0.95\hsize]{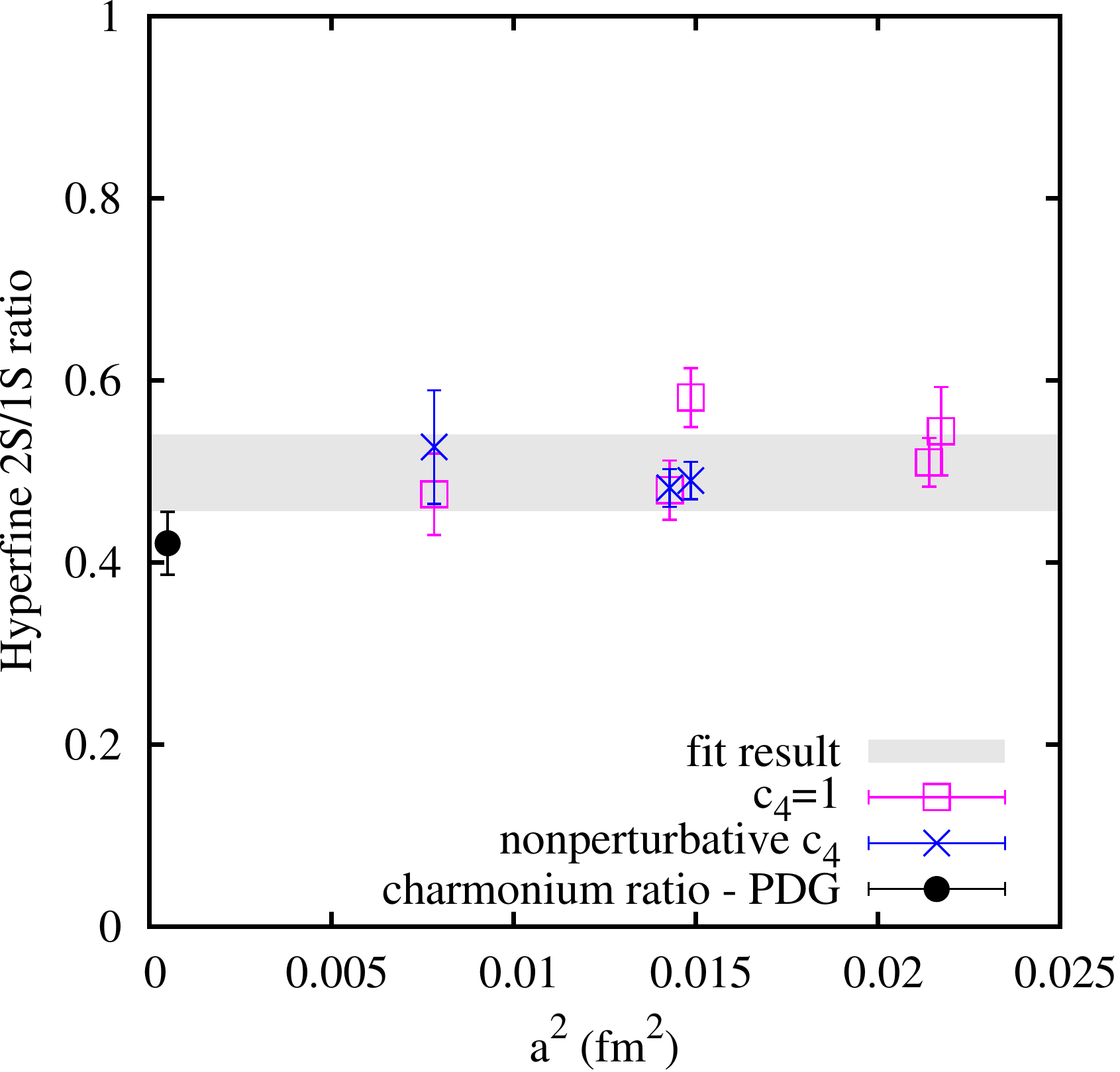}
   \caption{Results for the ratio of $2S$ to $1S$ hyperfine splittings from a $5\times5$ matrix fit. Pink open squares show the result for a tree level $c_4$ and blue crosses denote results using the nonperturbatively tuned values.
   The corresponding ratio for charmonium is included for comparison.}
   \label{fig:hyperfine2s}
  \end{minipage}
 \end{center}

\end{figure}

\paragraph{Hyperfine splittings:}
Another important test of any computation of the spectrum is the hyperfine splitting $M(\Upsilon) - M(\eta_b)$. 
Previous tree level lattice NRQCD calculations suffered from large systematic errors of around $25\%$. The dominant contributions to this error come from radiative corrections, since the splitting is proportional to $c_4^2$, and $\mathcal{O}(v^6)$ terms in the action. 
Figure \ref{fig:hyperfine} shows our new results for the hyperfine splitting. The plot shows the result from both perturbatively and nonperturbatively (using the experimental P-wave spin splittings) calculated $c_4$ values and the physical result extracted from this data. The values have been adjusted for mass mistuning and the effect of missing 4-quark operators has been calculated perturbatively. Correlated systematic errors for missing higher order terms in $\alpha_s$ were included in the fit.
Finally, a $10\%$ systematic error has been applied for missing $v^6$ terms.

The ratio of the $2S$ to $1S$ hyperfines is shown in figure \ref{fig:hyperfine2s}. This ratio should be independent of $c_4$, a fact which is confirmed by the data, so the error is dominated by statistics. 
Our results for the $1S$ and $2S$ hyperfine splittings are:
$$
M(\Upsilon(1S)) - M(\eta_b(1S)) = 70(9) {\rm MeV}, \ \ \ \ \ \ 
M(\Upsilon(2S)) - M(\eta_b(2S)) = 35(3) {\rm MeV}
$$

\paragraph{D-wave splittings:}
We are also able to expand upon previous NRQCD spectrum calculations by giving a prediction of the D-wave spin splittings. Several different lattice operators were used for each physical state and the result for the coarse ensemble is shown in figure \ref{fig:dwaves}.
The splittings have no noticeable lattice spacing or light quark mass dependence.

\begin{figure}
 \begin{center}
  \begin{minipage}[t]{0.48\hsize}
  \vspace{0pt}
   \includegraphics[width=0.95\hsize]{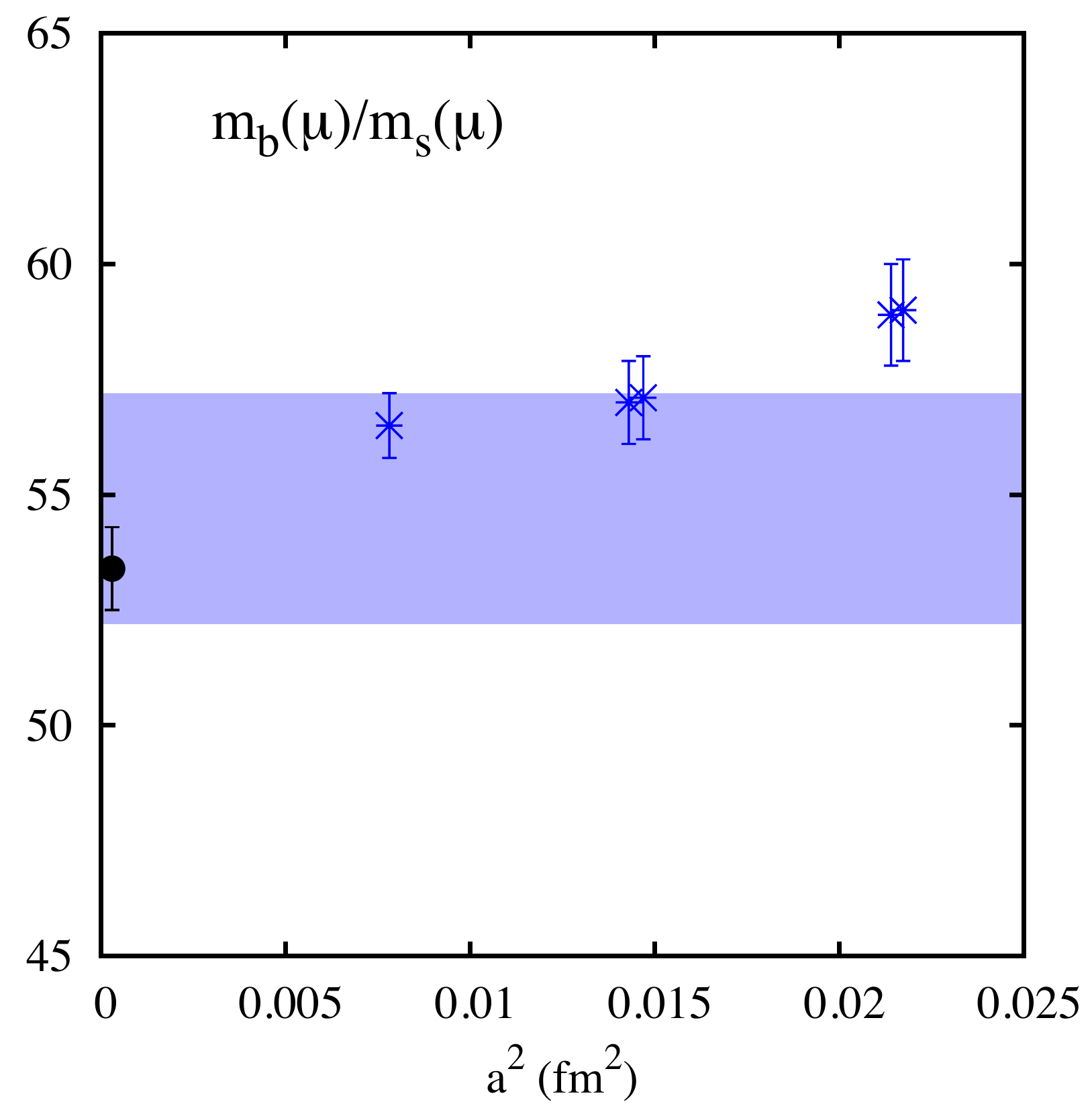}
   \caption{Plot of $m_b/m_s$ in the $\overline{MS}$ scheme at scale $\mu$. The errors on the points include statistical, lattice     spacing and NRQCD systematics.
    The previous HISQ result is shown with a black circle.}
   \label{fig:mbms}
  \end{minipage}
  \hfill
  \begin{minipage}[t]{0.48\hsize}
  \vspace{0pt}
   \includegraphics[width=0.95\hsize]{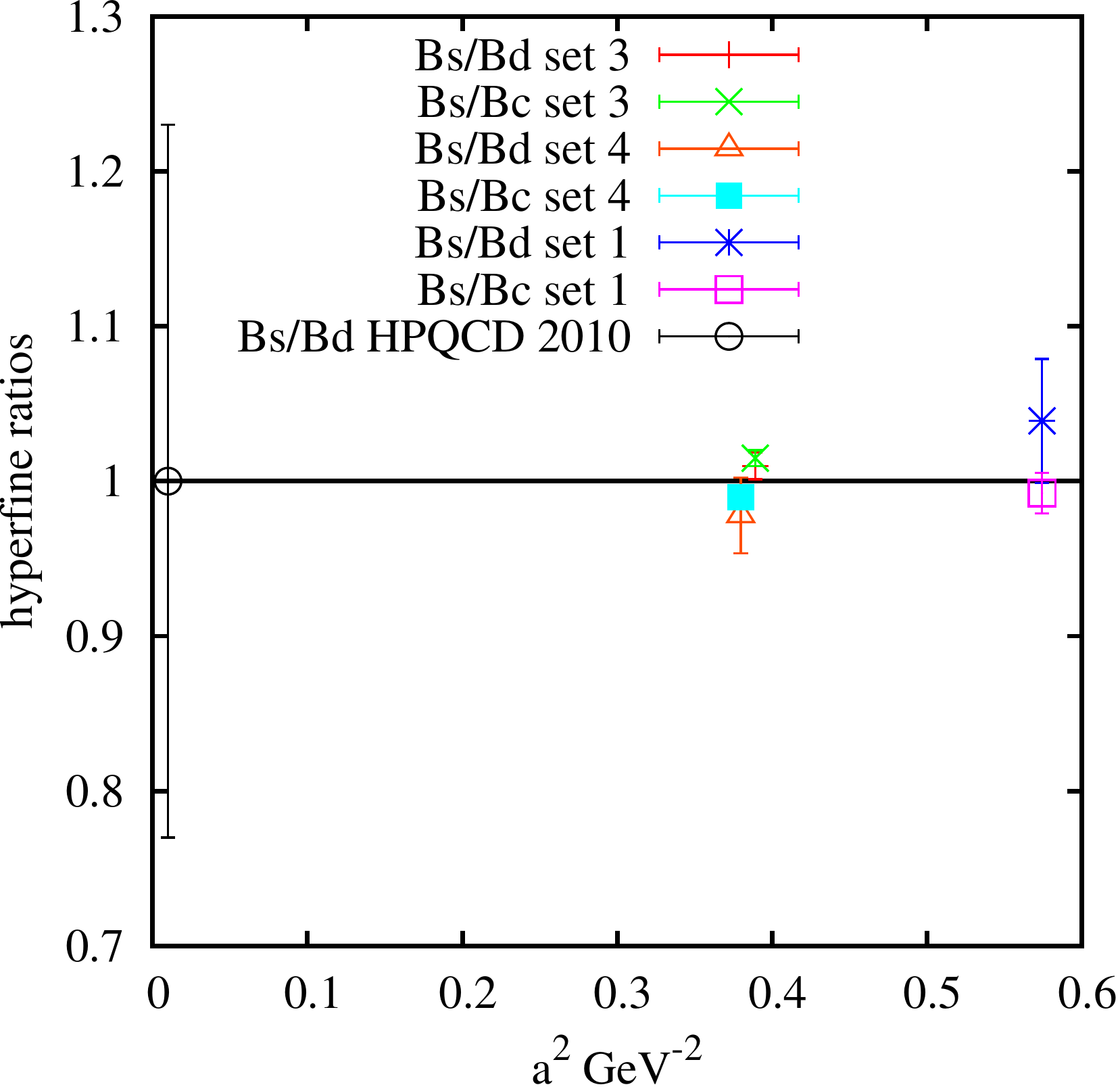}
   \caption{A plot of the ratio $(M_{B_s^*} - M_{B_s} ) / (M_{B_q^*} - M_{B_q}  )$ (where $q=d,c$) on two coarse (set 3,4, $a\sim 0.12$fm) and a very coarse (set 1, $a \sim 0.15$fm) ensemble. The previous HPQCD value of 1.00(23) is included for comparison.
    }
   \label{fig:bhyperfine}
  \end{minipage}
 \end{center}
 
\end{figure}

\paragraph{The ratio $m_b/m_s$:}
Using the tuned values of $am_b$ and $am_s$ we can calculate the ratio of the two quark masses. These are converted to the $\overline{MS}$ scheme at scale $\mu$ via the pole mass. The NRQCD and HISQ mass renormalisations are known to one loop and the dominant error comes from missing $\mathcal{O}(\alpha_s^2)$ terms. The result obtained is 54.7(2.5) and is shown in figure \ref{fig:mbms}. Also included is a previous, independent HPQCD value (53.4(9)) using HISQ for both the $b$ and $s$ quarks \cite{McNeile:2010ji,Davies:2009ih}.

\paragraph{B meson hyperfine splittings (preliminary):}
Here we discuss preliminary results for the low lying B meson states. 
The B meson masses are extracted from NRQCD-HISQ correlators with random noise sources using local and exponential source smearings with two different radii.
In figure \ref{fig:bhyperfine} we provide further evidence to support the previous HPQCD prediction in \cite{Gregory:2010gm} that ratios of B meson hyperfine splittings are equal to one - as above, the ratio should be independent of $c_4$. This implies that the heavy-light hyperfine splitting is independent of the light valence quark mass.

\section{Discussion}
The calculation of the bottomonium spectrum presented here shows a significant improvement in both systematic and statistical uncertainty over the previous HPQCD results with evidence for similar levels of improvement to come in the heavy-light sector. This is part of a broader heavy quark physics program in progress by the HPQCD collaboration and gives us the opportunity to accurately tune all necessary parameters and perform checks of the perturbative improvements.
The main aims of the program are improved determinations of important quantities for B physics phenomenology using the NRQCD and HISQ actions.
Our results demonstrate that our errors are well understood and that subsequent computations will be reliable.

\section*{Acknowledgements}
We are grateful to the MILC collaboration for the use
of their gauge configurations and values for $r_1/a$.
The results described here were obtained using the Darwin Supercomputer 
of the University of Cambridge High Performance 
Computing Service as part of the DiRAC facility jointly
funded by STFC, the Large Facilities Capital Fund of BIS 
and the Universities of Cambridge and Glasgow

\bibliographystyle{hieeetr}
\bibliography{lattice_bib_08-11}

\end{document}